\documentstyle[aps,epsf,rotate,prl]{revtex}
\tightenlines
\newcommand{\be}{\begin{equation}}
\newcommand{\ee}{\end{equation}}
\newcommand{\ba}{\begin{eqnarray}}
\newcommand{\ea}{\end{eqnarray}}

\draft        
\begin{document}             
\title{Comment on "Phonon Spectrum and Dynamical Stability of a Dilute Quantum Degenerate Bose-Fermi Mixture}
\maketitle




In a Letter, Pu et al \cite{Pu02} studied the phonon spectrum of 
a dilute Bose-Fermi Mixture.  They claimed that there
is a dynamical instability of the system for 
certain parameter regimes.  They based their conclusion on
an analysis of the dispersion relation of the phonons,
their eq (14), and conclude that
an instability arises if process shown in their Fig. 1
can occur.
Their claim is incorrect.  The process in their Fig. 1
represents in fact Landau {\it damping} \cite{LL} of
the collective mode but not an instablity of the system.
The process depicted over there {\it absorbs} energy from the 
sound wave.  (c.f., ref \cite{LL2})

  Their eq (14) has actually been derived and studied before 
\cite{Yip01}.  The effect of Landau damping on
the mode propagation has been analyzed over there.

\vskip 1 cm

\noindent
S. K. Yip $^{*}$\\

\noindent
Institute of Physics, \\
Academia Sinica, \\
Nankang, Taipei 11529, Taiwan 

\vskip 0.5 cm


\vskip 1 cm

\noindent $^{*}$ Email address: yip@phys.sinica.edu.tw

\end{document}